# Predicting the binding of small molecules to proteins through invariant representation of the molecular structure

*R. Beccaria[1,2,3], A. Lazzeri[1] and G. Tiana[1,4]\**

[1]Department of Physics, University of Milano, via Celoria 16, 20133 Milano, Italy.
[2]Heidelberg Institute for Theoretical Studies, Schloss-Wolfsbrunnenweg 35, 69118 Heidelberg, Germany
[3]Faculty of Physics, Heidelberg University, Im Neuenheimer Feld 227, 69120 Heidelberg, Germany
[4]INFN, via Celoria 16, 20133 Milano, Italy.

**Abstract**
We present a computational scheme for predicting the ligands that bind to a pocket of known structure. It is based on the generation of a general abstract representation of the molecules, which is invariant to rotations, translations and permutations of atoms, and has some degree of isometry with the space of conformations. We use these representations to train a non-deep machine learning algorithm to classify the binding between pockets and molecule pairs, and show that this approach has a better generalization capability than existing methods.

**Introduction**
Predicting the ability of a small molecule to bind to a pocket in a protein is the first step in computer-assisted drug design. Structure-based algorithms are becoming increasingly popular for this purpose, especially after the development of tools to predict the three-dimensional structure of protein targets using only genetic information.

All ligand scoring methods are partially or fully tuned on empirical data, that is making use of databases of experimentally determined pocket-ligand pairs. Some of them, such as Autodock [1] or Vina [2], make use of energy functions whose shape is based on the physical interaction between the classically described atoms (such as Lennard-Jones, Coulomb functions, etc.) and whose parameters are determined empirically or semi-empirically. They work exploiting the idea that the more realistic the description of the interactions, the better the prediction of the binding properties. However, these energy functions are only approximations to the true molecular interactions, which depend on the quantum wave function of the system and on the environment, and their performance is not always optimal [3].

Machine learning methods are becoming increasingly interesting for the assessment of protein-small molecule interactions. In particular, the use of multi-layer artificial neural networks can approximate any functional form of the true interaction energy, including many-body and entropy-based effects, as guaranteed by the Universal Approximation Theorem [4]. Thus, their use makes the interaction model independent of the choice of a specific energy function.

For example, convolutional neural networks similar to algorithms used in image recognition, have been applied to a grid representation of the structure of proteins and ligands and were successfully trained to classify binders and decoys [5]. Graph neural networks (GNN), which take graphs made of nodes and edges as input, are particularly well suited to describing molecules in the classical approximation. Several types of GNN have been developed to predict the binding affinity between proteins and ligands. SIGN [6] describes the atomic environment of each atom in polar coordinates. In PigNet [7] the nodes are described to interact through realistic physical functions (e.g. Lennard-Jones functions, etc.). In both GraphCNN [8] and DeepBindGCN [9] the properties of the protein pocket at residue resolution and the ligand are first converted into an abstract representation, which is used as input for a fully connected network. In GraphCNN the node properties are 80 physicochemical features of the residues for the pocket and the atom type for the ligand, and the abstract representation is set before training the classifier. In DeepBindGCN, the properties of the nodes are the residue type for the pocket and the Smile code for the ligand, and the molecular representation is obtained during training of the whole classifier.

---

\* Correspondence to guido.tiana@unimi.it

The search for a significant abstract representation of molecules is itself a goal that was pursued with GNN. A graph convolutional neural network was used to encode 13 physicochemical features of the atoms that make up small molecules into a representation capable of predicting properties of the molecule such as solubility and melting point [10]. An efficient representation can be used for ligand comparison or for searching based on molecular similarity.

A key feature of machine-learning algorithms for classifying or quantifying the binding between the pocket of a protein and a small molecule is that their prediction must be invariant for translations and rotations of the atomic coordinates and to permutations of the indices of the atoms [11]. GNNs usually depend on interatomic distances and then their output is automatically invariant for translations and rotations, but not for permutations. They are equivariant for permutations, meaning that a permutation of the order of the atoms in the input structure produces a corresponding permutation in the representation. A function of the representation can be invariant if it contains a pooling operation that sums the contribution of all nodes, as in the case of physics-informed functions that predict the dissociation energy of the ligand, in the form of a sum of equivariant terms [7]. In any case, the intermediate representation of the molecules is usually not invariant for permutations.

In the present work, we first build an invariant universal representation of the pocket and of the ligand, that is a fingerprint based only on the atomic coordinates and independent of their mutual binding properties. We then use these representations to train some non-deep machine learning algorithm to classify the binding between pockets and ligands, and show that this approach has a better generalisation capability than existing methods. The choice of non-deep algorithms is a consequence of the scarcity of protein-ligand structures, which we believe discourages the use of the deep-learning methods reported in the literature and described above.

**Methods**

The strategy we employed is made of two disjoint steps (Fig. 1). First, we use a graph autoencoder to create an abstract representation of the pockets and of the ligands, which is invariant for rotations, translations and permutations of the atoms, taking advantage of the large amount of available (non-complexed) structures. We then use a random-forest algorithm to classify binding against non-binding pairs of pockets and ligands, described by their abstract representation.

*The datasets*

Three datasets of protein-ligand structures are considered in the present work. For each protein, we selected the 30 heavy atoms closest to the ligand and regarded this set of atoms as the binding pocket. Analogously, we only selected ligands with a maximum of 30 atoms. The three datasets are

The DUD-E dataset [12] consists of 102 target proteins associated to an average of 224 binding compounds and 50 decoys per target. Binding compounds are obtained from crystallographic structures, while decoys are ligands selected to have similar physicochemical properties as binding compounds but dissimilar topology; thus, positives are experimentally validated while negatives are not.

The MUV dataset [13] consists of 17 proteins, each binding and average of 30 ligands and associated with $10^3$ decoys. Here, both positives and negatives are validated experimentally.

In the PDBbind dataset [14], $10^6$ protein-ligand structures are extracted from the pdb and associated with their experimentally measured binding affinities. These include redundant pockets and any kind of small molecule (including non-druggable ones such as, e.g., acetylcholine, citric acid, propylphosponic acid). No decoys are reported. We extracted 436 structures with the requirement that there were no repetitions in the selected complexes, implemented simply comparing the protein names and the ligand identifier.

*Graph autoencoder*

The encoding part is made of an equivariant graph encoder, implemented with the e3nn PyTorch library [15]. The nodes of the graph describe the atoms of the molecule and their features $f_i$ are initialized with a one-hot encoding of the atom type. Two nodes are connected if their Euclidean distance, obtained from the crystallographic coordinates of the atoms, is smaller than 10Å.

A continuous filter convolution is applied iteratively to the nodes, namely

$$f_i^{t+1} = \frac{1}{\sqrt{z}} \sum_{j \in \partial(i)} f_j^t \otimes \left(h(r_{ij})\right) Y_m^l \left(\frac{\vec{r_{ij}}}{r_{ij}}\right), \quad (1)$$

where $z$ is the average degree of the nodes, $\partial(i)$ is the set of the nodes connected to the $i$th, $\vec{r_{ij}}$ is the vector of the distance between nodes $i$ and $j$, $r_{ij}$ is its norm and $Y_m^l$ are spherical harmonics, which are applied to the direction of the link. The operation $x \otimes (w) y$ is a tensor product of $x$ with $y$ parametrized by some weights $w$. The map $h$ is a 3-layers perceptron where each layer has 100 fully connected neurons, while distances are embedded using a basis of 10 cosine functions. In our graph encoder, we performed four convolutions. The output of the encoder is then a latent vector $f_i$ whose dimension is the number of atoms of the molecule and that is invariant with respect to roto-translations of the atomic coordinates and with equivariant with respect to permutation of the indexes of the atoms (cf. Extended Methods in the Supp. Mat.).

The decoder is a fully connected feed-forward network that takes as input the latent vector and reconstructs the experimental distance map of the atoms of molecule. To deal with the different size of the molecules in the latent space, a padding layer is used at the input layer of the network (more information in Supp. Mat.). The loss function that we minimize is the mean square error (MSE) between the predicted and the experimental distance map between the pairs of atoms $i$ and $j$

$$\mathcal{L} = \sum_{i<j}^{N} \left(d_{ij} - d_{ij}^{exp}\right)^2 . \quad (2)$$

The two autoencoders for pockets and ligands were separately trained with a stochastic gradient descent using Adam [16] as optimizer and an early stopping was set to save the model which performs the best on the validation set. Several hyper-parameters were tested leading to an optimal model on the validation set in which a learning rate of 0.001, 3 decoder layers each consisting of 500 fully connected nodes and a batch size of 128 for the ligands and of 8 for the pockets were used.

The permutation-equivariant (also invariant for translations and rotations) latent vectors undergo a pooling procedure to produce a permutation-invariant fingerprint representation of the molecules (Fig. 2). The pooling procedure consist of generating a histogram of the $f_i$ separately for each atom type. The range for each atom type in the histogram is selected to contain the 95 per cent of the latent values of that specific atom and two additional bins at the edge of the histograms are added to account for outliers. The matrix $M$ obtained by combining together the histograms is the abstract representation that we will use as the fingerprint of the molecule.

*Random forest*

We made use of the random-forest classifier [17] of scikit-learn to distinguish which pairs of protein pockets and ligands can bind to each other from those that cannot. The random forest with the best performance was found to be the one with 500 trees each with a maximum depth of 100 levels and using the entropy as a criterion for evaluating splits.

*Feed-forward network to predict physico-chemical properties*

We studied the correlation between the encoding matrices of the ligands with their physico-chemical properties (e.g. polar area surface mass) in different ways. First, we tried a linear regression between the norm

$$\lambda_M = \left(\sum_{ij} M_{ij}^2 B_{ij}^2\right)^{1/2} \tag{3}$$

of the encoding matrix $M$, where $B$ is the matrix containing the values associated to the bins for the different atom types, and the numerical value associated with the different properties, as obtained from PubChem [18], also trying semilog and log-log transformations. For highlighting more subtle correlations, we used a shallow feed-forward fully-connected network with the molecule matrix representation as input, or 2 hidden layers each with 10 fully connected layers and a linear activation function to predicted, and the numerical value of the physico-chemical property as output.

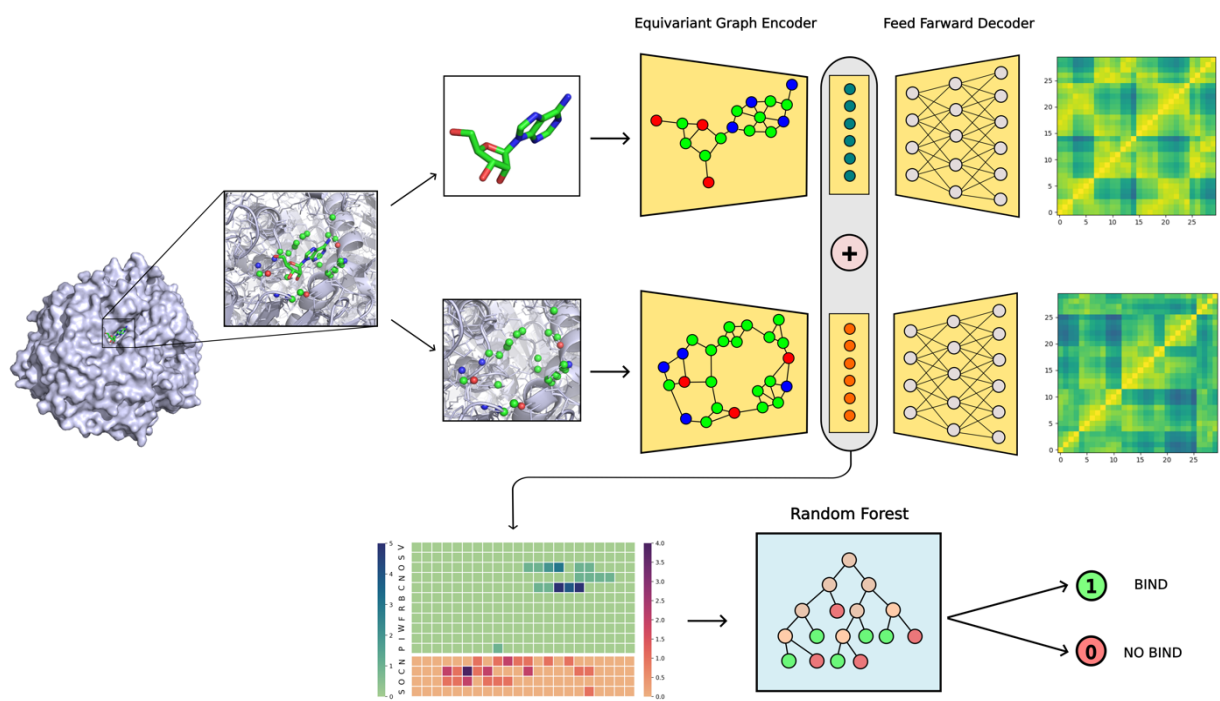

Figure 1: The model is built of two modules, an equivariant autoencoder (upper part) to produce a compact representation of the molecular structure in terms of a matrix (lower-left part), and a supervised non-deep algorithm to predict the binding of a ligand to a given pocket in the protein (lower-right part).

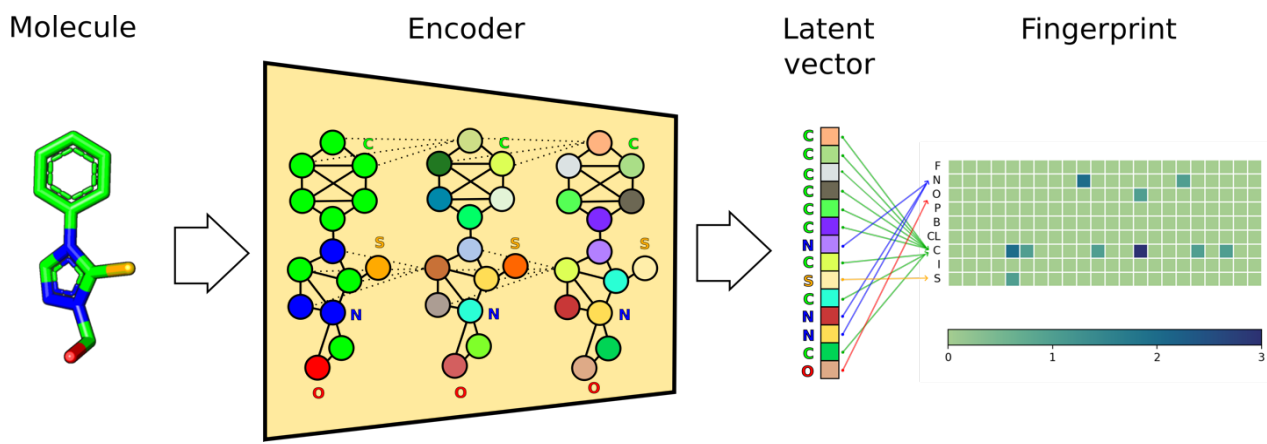

Figure 2: The abstract representation of a molecule (fingerprint) consists of a pooling operation of the latent vector obtained via the graph encoder. In the pooling operation, a histogram is generated for each atom type. The matrix resulting from this set of histograms is the representation of the molecule considered in the present work.

## Results

*The autoencoder produces a general representation of pockets and ligands*

We first verified the ability of the autoencoder to produce an abstract representation of pockets or ligands that faithfully describes them, in the sense that it is possible to recover from the representation the three-dimensional conformation of their atomic structure. Such a representation is constructively invariant with respect to rotation, translation, and permutation of the atoms. The matrix resulting from this encoding will be the fingerprint of each pocket and ligand.

We constructed a dataset for protein pockets by merging all the molecular structures obtained from the DUD-E, from the MUV and from the PDBbind databases; the resulting dataset contains 546 pockets. Analogously, we constructed a dataset of 37806 ligands, merging those obtained from DUD-E and MUV. We did not include ligands from PDBbind because their non-redundant set is negligibly small. We used 80% of each dataset for training and the remaining 20% as a test set.

The mean square error (MSE) of the training set typically reaches a low steady state in few hundreds training epochs (Fig. 3a). The average MSE is around 1Å for ligands, while is around 2Å for protein pockets. As a comparison, the median resolution of X-ray structures in the pdb is 2.05Å, indicating that we can encode the molecular structures with a resolution that is similar to the experimental one. The distribution of training and of validation errors are similar to each other both for pockets (Fig. 3b) and for ligands (Fig. 3c), suggesting that we are not overfitting the input dataset.

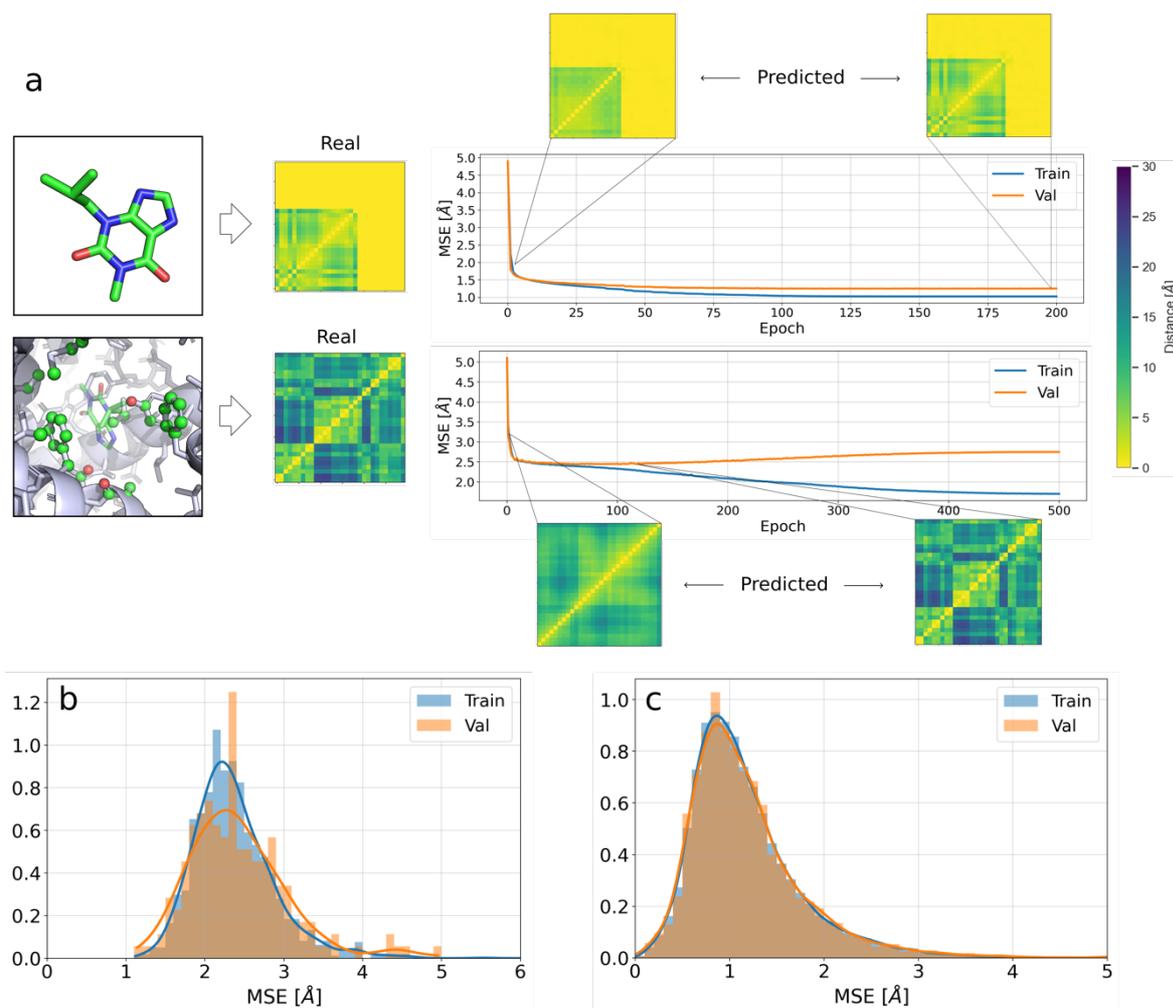

Figure 3: The MSE of the autoencoders between real and predicted distance maps for ligands and pockets during the training. a) the autoencoder is trained to reproduce the structure of the molecules structures in terms of their distance maps: the predicted distances maps improve as the epoch increases. b) distributions of the MSE values for pockets in the training and in the validation sets with their relative mean values. c) distributions of the MSE values for ligands in the training and in the validation sets with their relative mean values.

*The encoding matrices are nearly isometric to the molecular structure*

We then investigated whether Euclidean distances in the space of pocket conformations display any correlation with some distances in the encoded space. Since the two spaces have different dimensionality, a complete isometry is not possible. To verify the existence of correlations, at least at small distances, we performed a molecular-dynamics simulation of the kinesin spindle protein (pdb code 2FL2) with the Amber force field at 350K, recording 500 conformations. From them, we extracted the coordinates of the atoms defined to constitute the binding pocket in the crystallographic structure, calculated the associated dRMSD and generated the representation matrices of the simulated structures using the pre-trained autoencoder.

We used the Jensen-Shannon divergence $D_{JS}$ (Eq. S5 in the Supp. Mat.) as a metric to compare pairs of encoding matrices, as the encoding representation is formally a set of distributions and $D_{JS}$ is one of the standard distances used to compare distributions .

The correlation between dRMSD in the space of atomic conformations and $D_{JS}$ between the corresponding representations depends strongly on the number of bins (Fig. 4). The best result is a correlation coefficient of 0.48 obtained with 4 bins. Interestingly, generating alternative conformations simply by adding a Gaussian noise to the crystallographic structure, gives a lower correlation coefficient of 0.27 (Figs. S2 in the Supp. Mat.). This difference indicates that the encoding of realistic pockets displays better properties than that of chemically unachievable ones. On the other hand, variation of the types of atoms in the pockets does not lead to a clear correlation with variation in the encoding matrix (Fig. S3 in the Supp. Mat.); this could again be because the random variation of atom types at fixed coordinates produces physically unrealistic compounds.

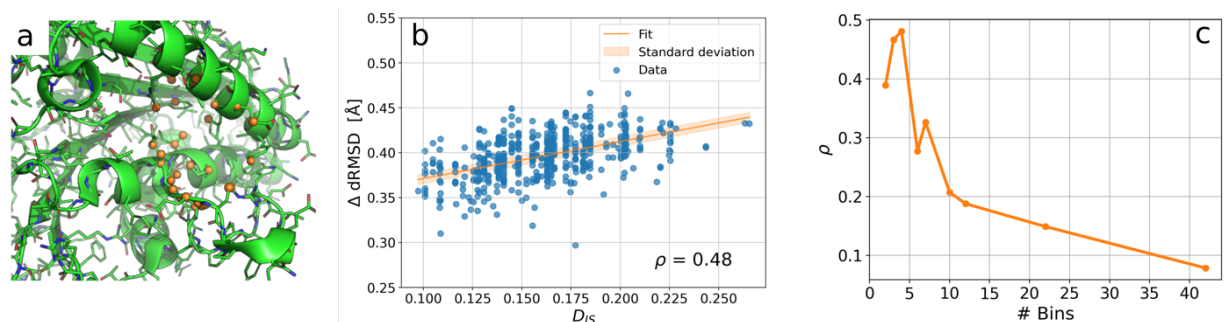

Figure 4: a) the 30 heavy atoms of of pocket of the protein kinase spindle protein (pdb code 2FL2, in orange). b) a scatter plot of the variation of the dRMSD of the pocket versus the variation of its representative matrix M upon changing its structure by molecular dynamics. c) the dependence of the correlation between dRMSD and $D_{JS}$ on the number of bins selected to construct the matrix from the latent vectors of the autoencoder.

*The encoding matrices are related to the physicochemical properties of the ligands*

Some physicochemical properties of the ligands can be obtained from PubChem (cf. Table S2 in the Supp. Mat.) and compared to the encoded latent vectors. The volume (V), the molecular weight (MW), and the molecular complexity [19] are strongly correlated with the norm $\lambda_M$ (see Methods) of the encoding matrix $M$ (Fig. 5). In particular, $\lambda_M$ is fitted by an exponential function of each of these quantities (coefficient of determination $r^2 = 0.54$ for the molecular weight, $r^2 = 0.76$ for the Volume and $r^2 = 0.40$ for the molecular complexity). An exponential relation $\lambda_M = A \exp[x/x_0]$ between the norm of the encoding matrix $M$ and a quantity $x$ indicates that there is an elementary scale $x_0$ that controls the variation of that quantity according to the encoding. This scale is 80.01 Å$^3$ for the volume, 107.05 g/mol for the molecular weight and 494 for the molecular complexity. None of the other properties reported in PubChem can be fitted with $\lambda_M$ and does not define straightforwardly an elementary scale (Figs. S4-S7 in the Supp. Mat).

To highlight more subtle relations between the encoded matrices and the physicochemical properties of the ligands without hypothesizing a specific functional form, we employed a shallow linear neural network, testing if it can be trained to predict the property from the encoded matrix. In the case of XlogP (a measure of hydrophobicity) and of the polar area surface, we constructed a dataset with 111441 ligands with tabulated values of those properties from PubChem. The 80% of the dataset is used as a training set and a 20% as a validation set. We trained two shallow neural networks to predict the polar surface area and the XlogP from the fingerprint of the ligands obtained with our trained graph autoencoder. The performance of the two networks is evaluated using the relative error of the validation set

$$\delta = \left|\frac{target - predicted}{target}\right|, \tag{5}$$

that is $\delta = 0.265$ for XlogP and $\delta = 0.117$ for the polar surface area.

Summing up, the fingerprint representation of the ligands reflects some of their physicochemical properties, either in a clear way (for MW, volume and complexity) or in a more subtle way (for hydrophobic properties).

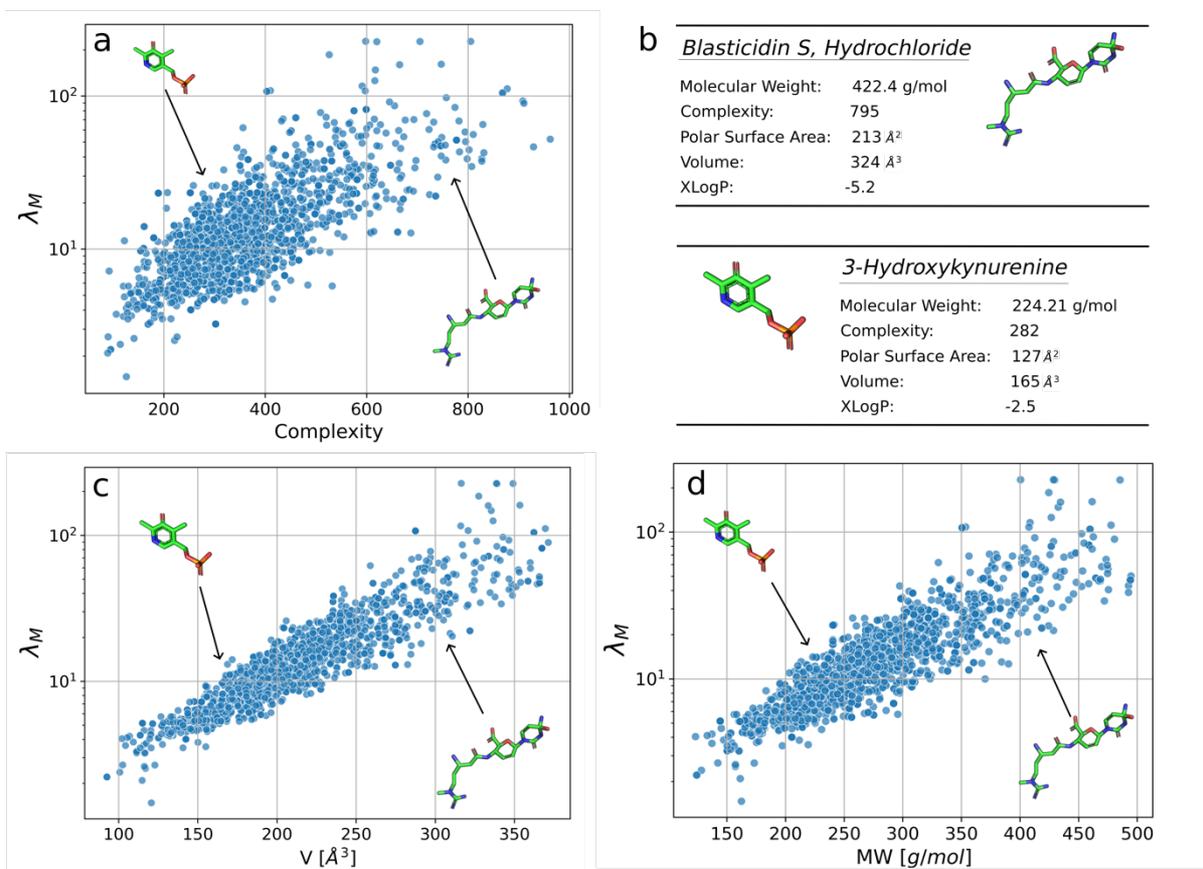

Figure 5: Correlations between chemical and structural properties of the ligands with the norm of their fingerprint. The high correlation with a power law manifests the presence of elementary scales encoded by the model.

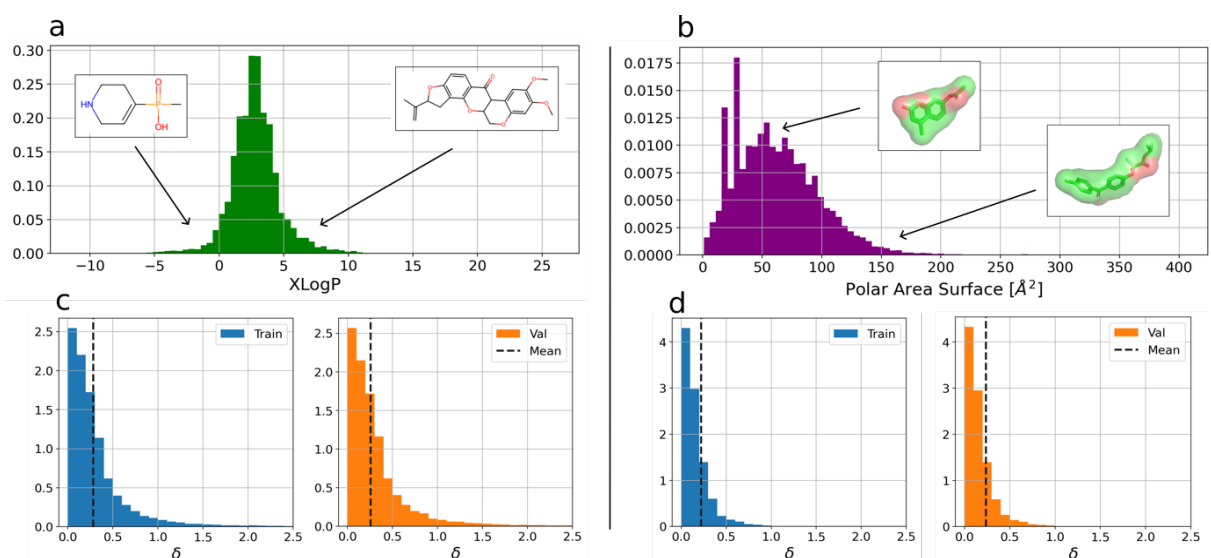

Figure 6: Predictions of chemical and structural properties of ligands from their fingerprints by a feedforward network. a) and b) distributions of XlogP and polar surface area values of the ligands as tabulated in PubChem. c) and d) distribution of the relative errors between real and predicted values of XlogP and polar surface area respectively.

*Binding ligands can be predicted from the encoded matrices*
The main reason to build invariant representations of the pockets and of the ligands in the form of encoded matrix is to employ them in the prediction of binding ligands for a given pocket. For this purpose, we made use of a random-forest classifier and tested its performance on the DUD-E and on the MUV datasets of ligands and decoys, training the algorithm on 80% of the dataset and using the remaining 20% as test set (Fig. 7). We removed the structures in the test set that are similar to those in the training set (Figs. S1 in the Supp. Mat.).
The training is nearly perfect for both datasets. The performance of the classifier on the test set is better on the DUD-E (with an AUC of 0.814) data than on the MUV data (with an AUC of 0.790). In particular, in the MUV dataset we detect slightly less true positives (69.05% vs 79.34%) and many more false negatives (30.95% vs 20.66%) than in the DUD-E dataset. The classification ability of the random forest is weakly dependent on the number of bins we use to build the encoding matrix (Fig. S8 in the Supp. Mat.).
The comparison of our random forest with other algorithms (Table 1) shows that we perform comparably to the best algorithms when both the training set and the test set are extracted from the DUD-E dataset. But to further challenge the generalization capability of our algorithm, we tested the algorithm trained on the DUD-E dataset also to the MUV and the PDBbind datasets. The reason to choose the DUD-E dataset as training set is that it is the largest.
For predicting the ligand binding of the MUV dataset, our random forest performs best among the prediction algorithms (Table 2), with an AUC of 0.671. Its generalization ability is then good also across datasets built independently from each other.
The PDBBind dataset does not contain any decoy, consequently we can only test what is the fraction of true positives we can identify. Overall, we have a true positive rate of 71%, that strongly depends on the size of the ligand (Figure 8). For small ligands it reaches 88%, while it decreases to 67% for ligands larger than 350 D, of which we retain only a subset of atoms, and which can be flexible and populate multiple conformations.
We also tested other non-deep machine learning alogirithm for classification, like supported vector machine, but random forest performs best (Fig. S9 in the Supp. Mat.).

| Algorithm | AUC |
|---|---|
| Random forest (this) | 0.814 |
| Graph CNNN | 0.886 |
| 3DCNN | 0.868 |
| Vina | 0.716 |
| RF-Score | 0.622 |
| NNScore | 0.584 |

Table 1: The performance of different algorithms on the DUD-E dataset, taken from [8].

| Algorithm | AUC |
|---|---|
| Random forest (this) | 0.671 |
| Graph CNNN | 0.621 |
| 3DCNN | 0.567 |
| Vina | 0.538 |
| RF-Score | 0.536 |
| NNScore | 0.454 |

Table 2: The performance of different algorithms trained on the DUD-E dataset and evaluated on the MUV dataset, taken from [8].

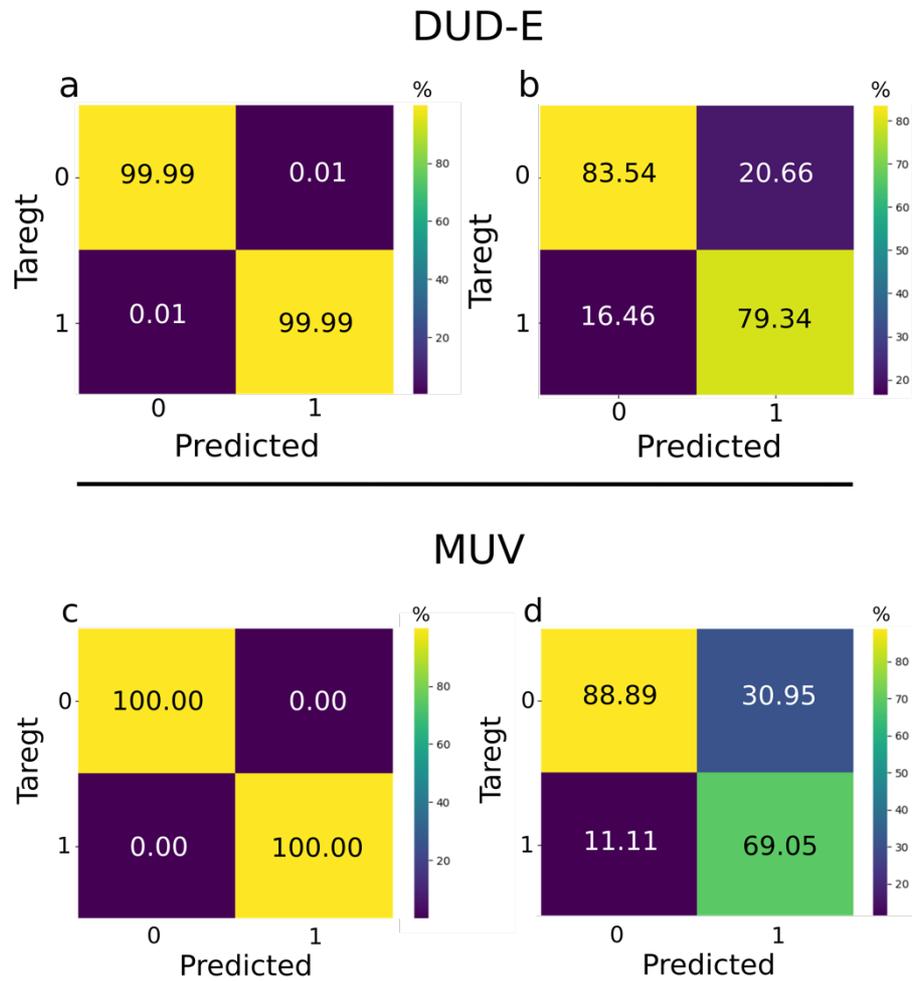

Figure 7: Confusion matrices obtained from two Random Forests separately trained and validated on the DUD-E dataset (a and b) and the MUV dataset (c and d). a and c are obtained from the training set, b and d from the test set.

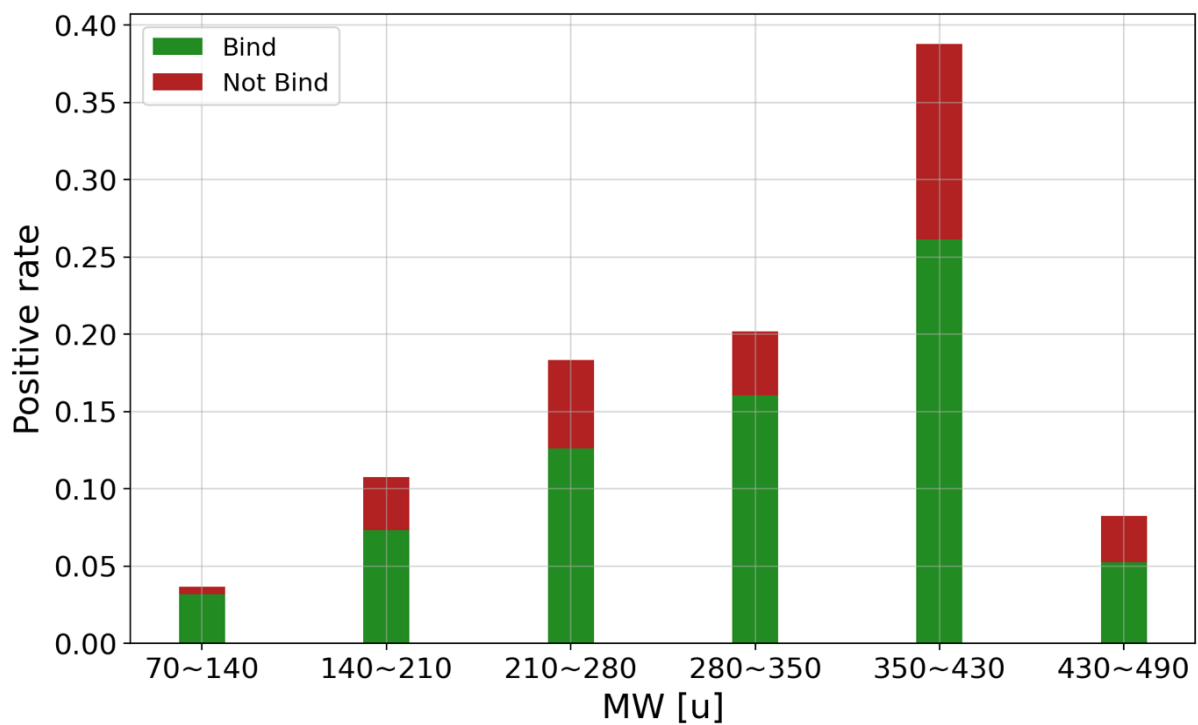

Figure 8: Fraction of true positive rates predicted by the Random Forest in the PDBbind dataset divided per weights of the ligands.

*Screening of ligands for a pocket*
We tested the encoding algorithm for a standard task of medicinal chemistry, that of virtually screening ligands in order to select a small set to test experimentally.
We used the pocket of the coagulation factor XII protease from the MUV dataset (target ID 852) as test case, whose crystallographic structure complexed with triacetyl-b-chitotriose is known (pdb 5exm). The screening is performed on all the ligands of the MUV dataset by representing both pockets and ligands with the encoding matrices obtained from the autoencoder and discriminating actives from decoys through the random forest, pretrained on the DUD-E dataset (Fig. 9). In the MUV dataset there are 1880 small molecules to test; among them there is triacetyl-b-chitotriose, which we regard as reference ligand because it is the one co-crystallized with the pocket selected for the test, and thus it is experimentally validated.

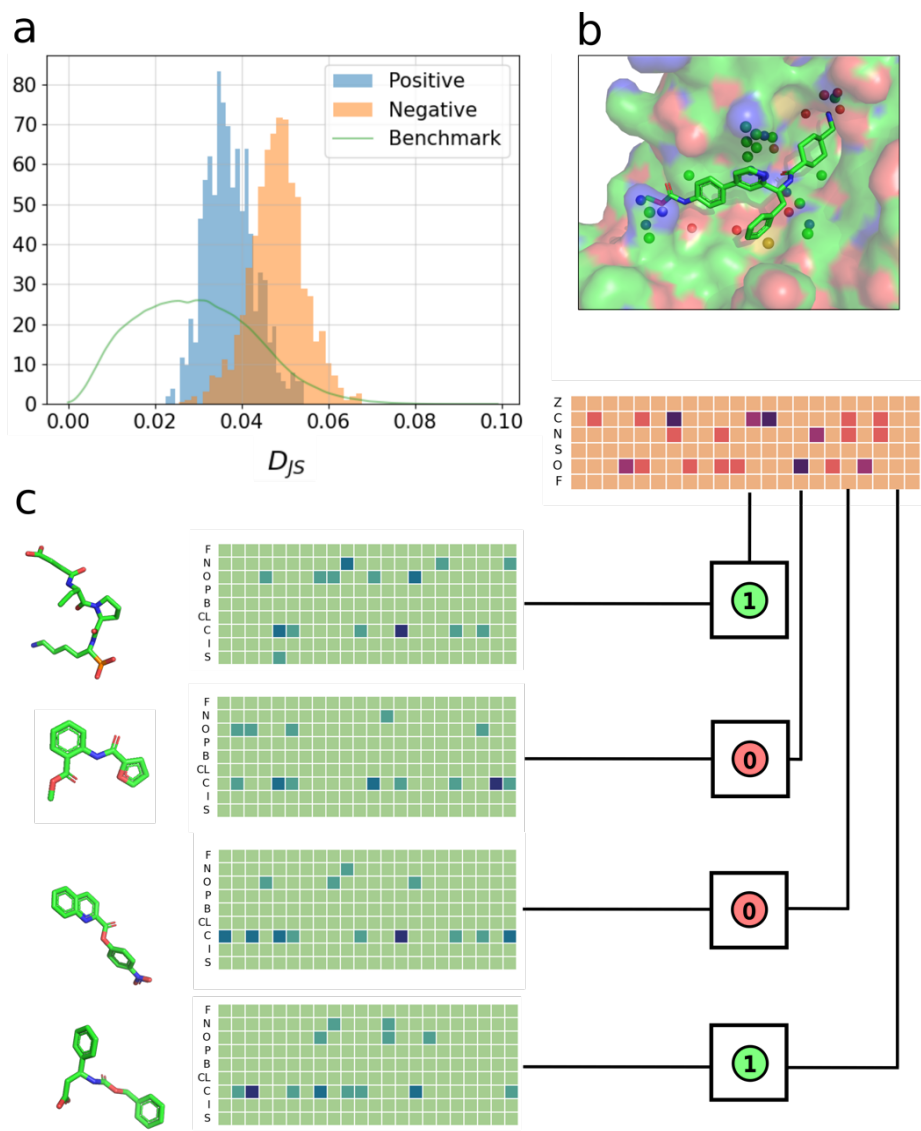

Figure 9: The screening of ligands to the coagulation factor XI protease (pdb code: 5exm), whose pocket binds triacetyl-b-chitotriose (b). Each ligand is encoded into a matrix by the autoencoder (green arrays) and classified as active ('1') or as a non-active decoy ('0') with a pre-trained random forest (c). The histograms (a) shows the Jensen-Shannon divergence between the representation of ligands and decoys to the experimentally-validated ligand triacetyl-b-chitotriose. The benchmark curve reports the mutual similarity of all ligands in the dataset.

The similarity of the encoding matrices of the ligands and of the decoys to that of triacetyl-b-chitotriose are quantified with the Jensen-Shannon divergence $D_{JS}$ (Fig. 9a). Ligands display a distribution of $D_{JS}$ which is significantly shifted to low values than that of decoys, indicating that the molecules predicted as ligands have a representation that is more similar to the experimental ligand than the predicted decoys.

**Discussion and Conclusions**

The scheme we have developed for predicting the ligands that bind to a pocket of known structure consists of two independent steps. First, we encode the pocket and each possible ligand in an abstract representation in the form of a matrix of tunable dimension. This matrix is invariant with respect to rotations, translations and permutations of the atoms of the molecule. It can be constructed to be nearly isometric to the conformation of the molecule, in the sense that the distances between close conformations correlate with the Jensen-Shannon distances in the space of the representative matrices, provided the conformations are chemically meaningful. The matrices encoded by the ligands also correlate with their molecular mass, volume and hydrophobic properties.

The decoder in the autoencoder requires the definition of a maximum number of atoms as input during the training, number which we chose to be 30 for computational efficiency. This number can be easily varied, and in any case does not limit the number of input atoms when the autoencoder is used for production, since the (encoding) graph neural network does not require the specification of an input dimension.

Classification algorithms are part of the basic arsenal for machine learning-based drug design [20]. The main problem one has to face to train such algorithms is the scarcity of patterns, due to the fact that crystallographic experiments are lengthy and challenging. As a matter of fact, the DUD-E database contains approximately 22,000 non-redundant complexes and the MUV database approximately 500. For this reason, we chose non-deep algorithms, which can be more efficient when training examples are insufficient [21]. The generalization error we obtained with random forest is indeed smaller than that got with other non-deep and deep algorithms.

The distinctive aspect of our computational scheme is the use of a structure-based representation of the pockets and of the ligands, which is generated independently of the classification from atomic-scale conformations. In this respect, our scheme is quite different from other machine learning approaches to ligand selection found in the literature. The approach that is most similar to ours is that of Graph-CNN [8], where an intermediate representation of pockets and ligand is generated. However, there the pockets are described at the level of amino acids, and the representation of pockets and ligands depends on their binding ability, as it is trained together with the binding classification (after an independent pretraining).

Our strategy has advantages and disadvantages compared to Graph-CNN. Among the formers is the generation of a purpose-independent representation of the atomic structure of pockets and ligands, that can be used for purposes that go beyond ligand design. Differently from other molecular fingerprints that encode only the chemical topology [22], our representation matrices reflect the three-dimensional atomic structure of the encoded system, which also turns out to correlate with some of its physico-chemical properties and to be approximately isometric to the conformational space. These properties can be useful in similarity searches, in QSAR, in QSPR, and also to develop tools to make the classification of ligands explainable.

Among the limitations of our atomic-resolution algorithms is the dependence of the results on the conformational fluctuations of the system, which is particularly relevant for flexible ligands and for cryptic pockets. Certainly, the isometric properties we found are expected to mitigate these problems and molecular dynamics simulations (e.g. the approach of [23]) can be used to generate a representative ensemble of conformations. Furthermore, our algorithm is not suitable for substructure search, since the matrices encoding a portion of a molecule are not guaranteed to be similar. Anyway, the good predictions we have obtained in classifying binding ligands suggest that these limitations are not overwhelming.

**Author contribution**
The manuscript was written through contributions of all authors. All authors have given approval to the final version of the manuscript.

**Associated content**
Extended methods and supporting figures in ligands-suppmat.pdf

The codes used to reproduce the results can be freely obtained from https://github.com/guidotiana/Milbinding and the parameters of the networks from https://dataverse.unimi.it/dataset.xhtml?persistentId=doi:10.13130/RD_UNIMI/5879ZG

# Supplementary Materials

**Extended Methods**

The algorithm we employed involves two steps, namely the generation of a molecular representation of a group of atoms (the pocket or the ligand) and the classification of pairs of representation of pockets and ligands into a binary outcome (binding or not binding).

First, we build an equivariant representation of the molecule in a lower dimensional through an Autoencoder. The data extracted from the datasets are 30×4 arrays containing the Cartesian coordinates and a one-hot-encoding of the atom types (cf. Table S1) of the 30 atoms of the pockets and of the ligand.

The equivariance property is an extension of the invariant property. Given any map $f: \mathbb{R}^n \to \mathbb{R}^d$, it is called equivariant if the outputs of transformed input are transformed accordingly to the same symmetry group in the output space. Given a mathematical group $G$ defined by a set of transformations $g \in G$, the equivariant concept can be formalized by

$$f(D(g)\vec{x}) = D'(g)f(\vec{x}), \qquad (S1)$$

where $D(g)\vec{x}$ is the representation of the transformation $g \in G$ acting in the $\mathbb{R}^n$ space, while $D'(g)$ is the representation of the same transformation but in the $\mathbb{R}^d$ space.

The map into the latent space defined by the Autoencoder is equivariant in the Euclidean Group (i.e. the group defined by rotations, translations and reflections) and by permutations of the atoms of the molecule.

The Autoencoder consists of two components:

1. An equivariant graph encoder which embeds the molecule in the latent space.
2. A fully connected decoder which reconstructs the geometry of the molecule (i.e. the distance map of the atoms) from the latent vector.

The equivariant graph encoder is implemented with the e3nn PyTorch library. The 30×4 arrays containing the Cartesian coordinates and a one-hot-encoding of the atom types are given as input to the encoder: it builds a graph where each node represents an atom and where two nodes are connected if the two respective atoms are withing 10Å. Each node $i$, is initialized with a node feature $f_i^0$ equal to the one-hot-encoding of the atom types and in the layers of the graph a continuous convolution filter is applied to all the nodes of the graph (see Eq. (1) in the main text).

As described by Eq. (1), the multy-layers perceptron $h$ provides the weights for the tensor product operation. It receives the distance between two nodes as input where the distance is embedded using a basis of functions, i.e. the distance $r_{ij}$ is represented as the linear combination of some set of functions. The weights of the linear combination are the inputs for $h$. The e3nn PyTorch library allows the user to choose between different basis: Gaussian, cosine, Fourier and Bessel. In the present model, we choose a cosine basis functions with 50 functions. The activation functions used by $h$ are SiLU functions, i.e.

$$silu(x) = \frac{x}{1+\exp(-x)} \tag{S2}$$

The output of the graph encoder is a latent vector with a single scalar value for each atom of the molecule.

The fully connected decoder takes the latent vector as input and reconstructs the geometry (i.e. the distance matrix) of the molecule. We used the PyTorch library to implement the fully connected decoder. The decoder has three hidden layers each with 500 nodes and the activation function used is the ELU activation function,

$$ELU(x) = \{x \; if \; x > 0; \; \alpha \cdot (\exp(x) - 1) \; if \; x \leq 0 \}. \tag{S3}$$

Only the triangular half of the symmetric distance matrix is used as output. Since the number of input and output nodes for the decoder must be specified in advance, we have used a padding procedure that acts on the latent vector to deal with ligands whose number of atoms is lower than 30. Since the latent vector always has a dimension equal to the number of atoms of the corresponding ligand, the padding procedure consists of adding as many zeros to the latent vector as necessary so that the latent dimension of the vector is equal to 30, which is the dimension of the input of the fully connected decoder.

Then, we build a final representation of the molecule combining its latent vector together with its atom types. The final fingerprint is a sparse matrix $M$ where each row $M_i$ is a histogram of the latent scalar values associated with the $i$th atom type (cf. Figure 2 in the main text). Important hyperparameters of the algorithm are then the range and the number of bins to construct $M$ from the latent vectors.

For some atomic systems we observe a few outlayers for each type of atom, that is elements of the latent vectors whose order of magnitude varies very significantly from the mean value of the distributions. To select the range of the histograms of each atom type, we selected the range which contains the 95% of the values. Two bins were added at each end of the histograms to collect the outliers in both directions.

*Jensen-Shannon divergence*

The metric we used to compare pairs of encoding matrices is the Jensen-Shannon divergence

$$D_{JS}(M^{(1)}, M^{(2)}) = \sum_{i=1}^{n_{bin}} \sum_{j=1}^{n_{at}} M_{ij}^{(1)} \cdot \log\left(\frac{M_{ij}^{(1)}}{M_{ij}^{(2)}}\right) + M_{ij}^{(2)} \cdot \log\left(\frac{M_{ij}^{(2)}}{M_{ij}^{(1)}}\right), \quad (S4)$$

where $n_{bin}$ is the number of bins used to build the enconding matrices from the latent vector and $n_{at}$ is the number of atom types, is usually defined for strictly positive distributions. However, in our matrix representations of the molecules we also have zero entries. A standard approach to deal with this scenario is to smooth the distribution by using a Dirichlet prior: each entry is treated as a fraction $\frac{M_{ij}^{(1)}}{N_1}$ where $N_1 = \sum_{i=1}^{n_{bin}} \sum_{j=1}^{n_{at}} M_{ij}^{(1)}$ and by replacing these fractions with $\frac{M_{ij}^{(1)}+1}{N_1+L}$ with $L = \sum_{i=1}^{n_{bin}} \sum_{j=1}^{n_{at}} 1$ (i.e $L$ is the lenght of the flattened matrix). This results in calculatin the Jensen-Shannon divergence as

$$D_{JS}(M^{(1)}, M^{(2)}) = \sum_{i=1}^{n_{bin}} \sum_{j=1}^{n_{at}} \frac{M_{ij}^{(1)}+1}{N_1+L} \cdot \log\left(\frac{M_{ij}^{(1)}+1}{M_{ij}^{(2)}+1} \cdot \frac{N_2+L}{N_1+L}\right) + \frac{M_{ij}^{(2)}+1}{N_2+L} \cdot \log\left(\frac{M_{ij}^{(2)}+1}{M_{ij}^{(1)}+1} \cdot \frac{N_1+L}{N_2+L}\right).$$

(S5)

**Supplementary Tables and Figures**

| Ligand | Pocket |
|--------|--------|
| C | C |
| N | N |
| S | S |
| O | O |
| F | Zn |
| P | Fe |
| Br | |
| Cl | |
| I | |

Table S1: the atom types used for pockets and for ligands. When the pocket contains ions in the crystallographic structures, they are included in the definition of the pocket.

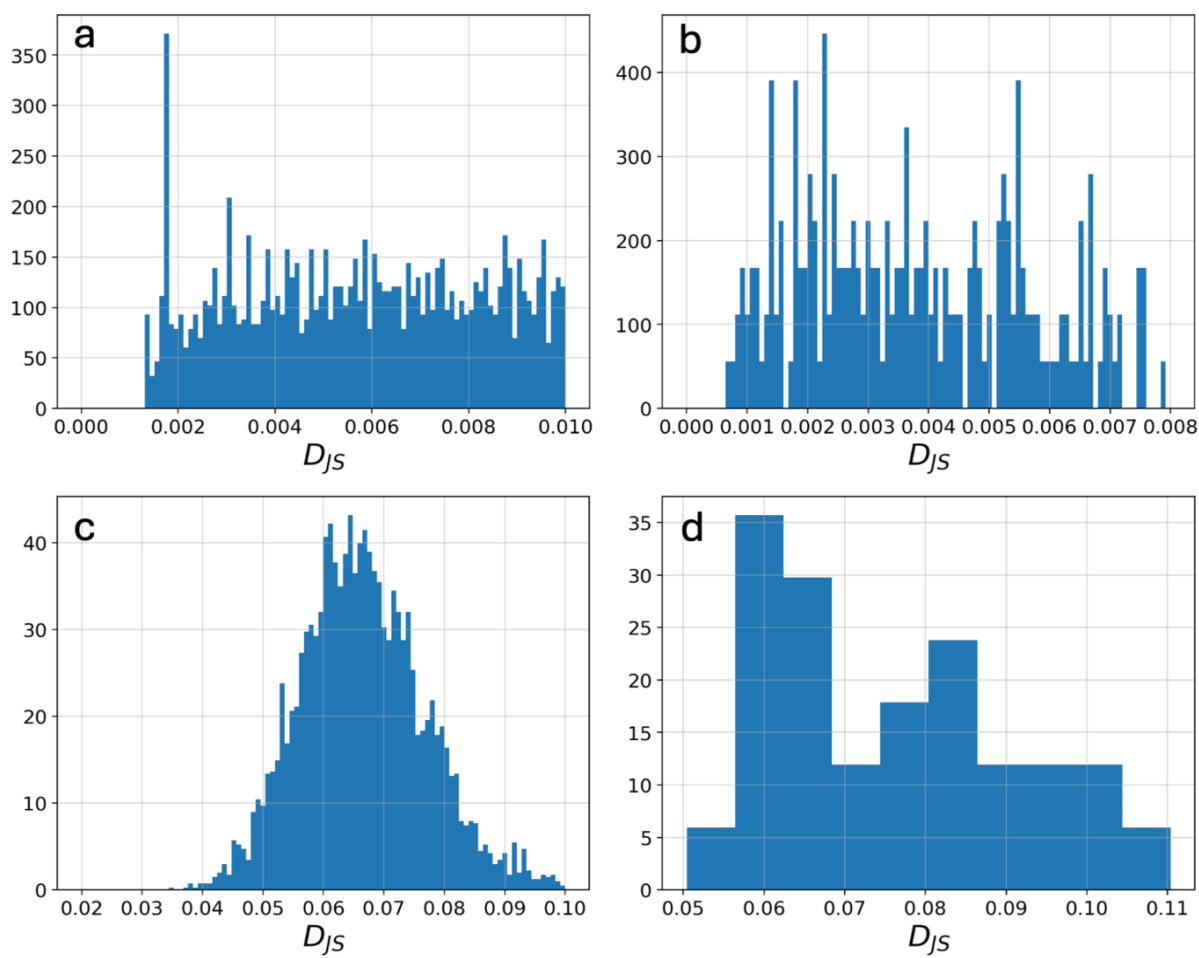

Figure S1: We checked that the test dataset is independent of the training dataset, discarding items whose representative matrix displays a distance $D_{JS}$ smaller than 0.03 for the pockets and smaller than 0.0007 for the ligands. a) and b) shows the $D_{JS}$ distibutions between ligands in the validation set and ligands in the training dataset for the DUD-E and MUV respectively after the selection. c) and d) shows the $D_{JS}$ distibutions between ligands in the validation set and ligands in the training dataset for the DUD-E and MUV respectively after the selection.

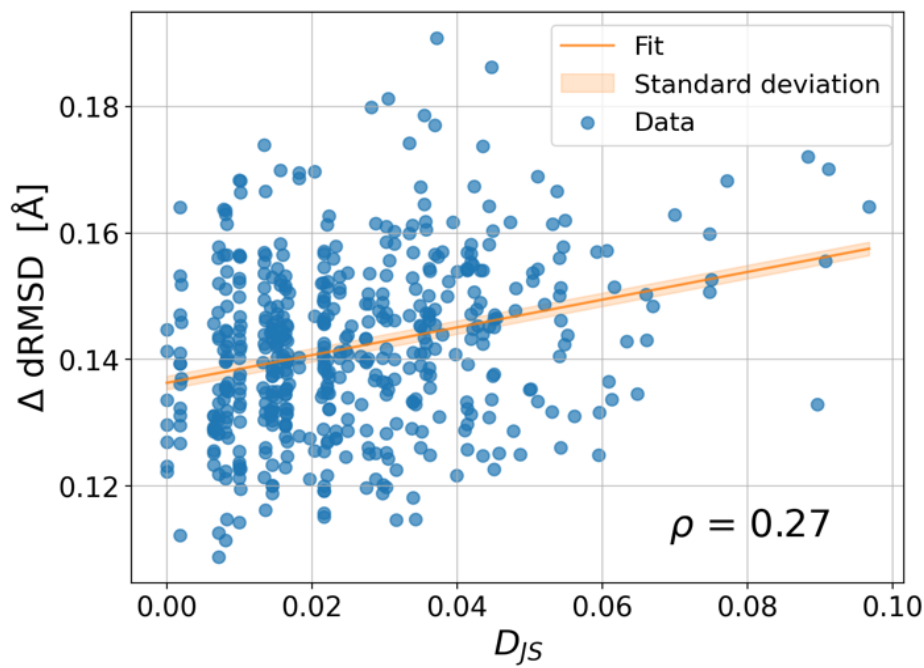

Figure S2: we generated 500 conformations by adding a Gaussian noise (1Å standard deviation) to the pocket of the Kinesin Spindle protein (pdb 2FL2). We correlate the dRMSD of the generated structures with the $D_{JS}$ of the respective encoded matrixes obtaining a correlation coefficient of 0.27. Interestingly, it is smaller compared to the correlation coefficient obtained using the configurations generated via the MD simulations.

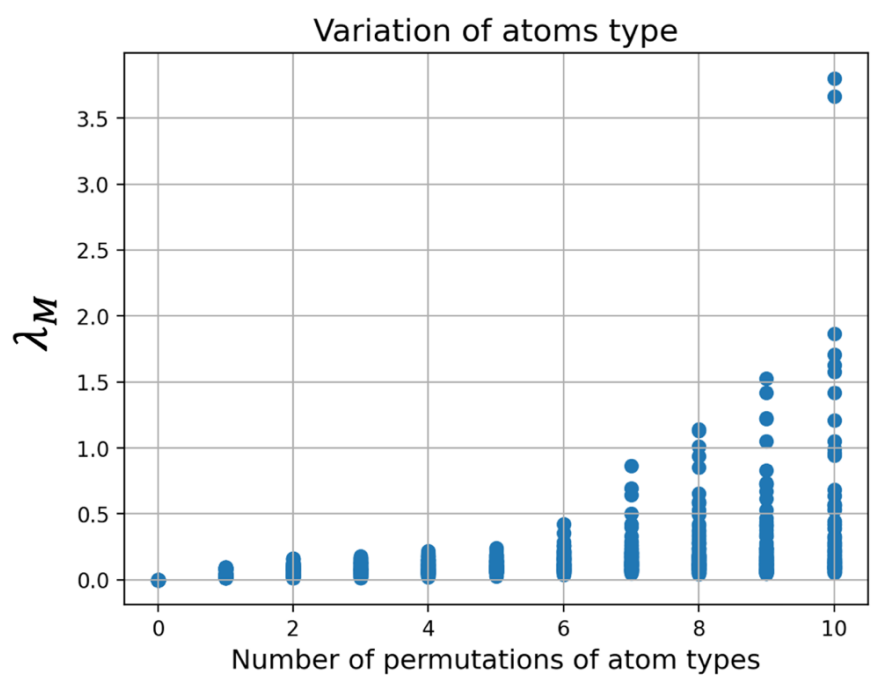

Figure S3: the permutations of the atom types in the pocket structure of the Kinesin Spindle protein does not correlate with the variation of the norm in the encoding matrix.

| Molecular weight | Volume |
| --- | --- |
| Complexity | XLogP |
| Polar area surface | BondStereoCount |
| HBDonorCount | HBAcceptorCount |
| RotableBondCount | |

Table S2: the physico-chemical properties of the ligands obtained from PubChem that we tried to correlate with the molecular encondings.

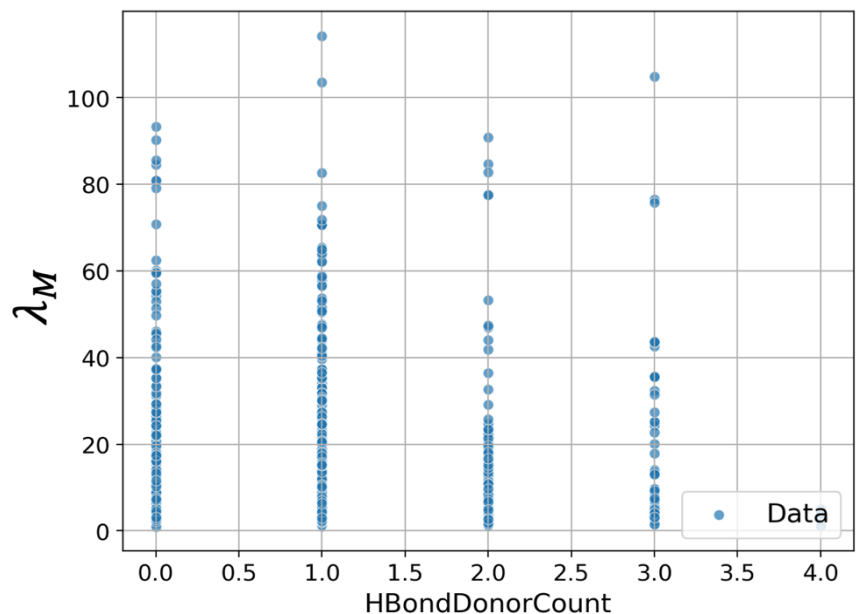

Figure S4: in PubChem each ligand is associated to the number of H-bond donor atoms. Here we compare this quantity together with the norm $\lambda_M$ values from the encoded matrix, for different ligands.

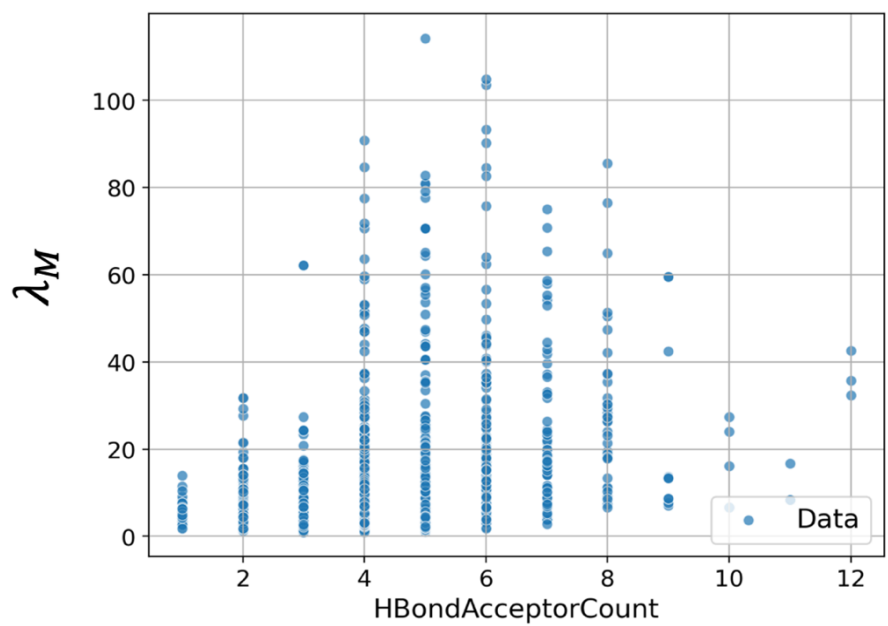

Figure S5: the scatter plot of the number of H-bond acceptor atoms against the norm $\lambda_M$ of the encoded matrix, for different ligands.

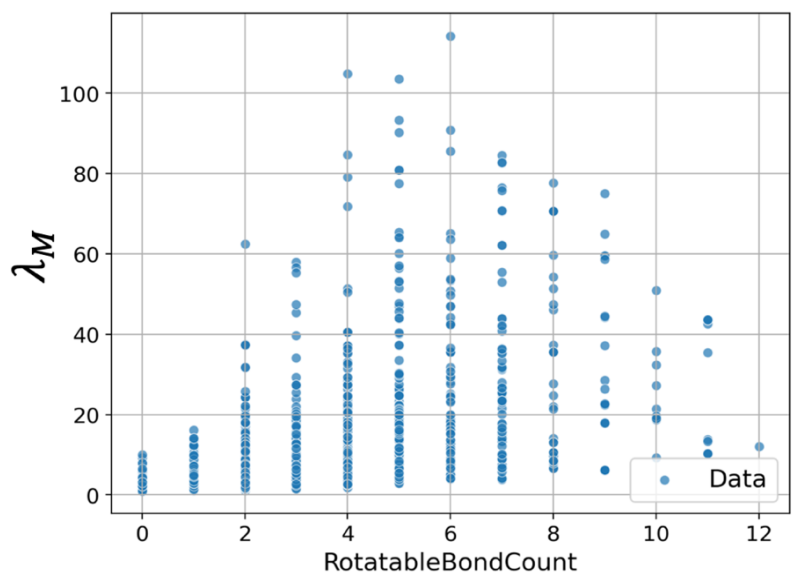

Figure S6: The number of rotable bonds is the number of bonds for which there is a free rotation around them. Here we plot this quantity together with the values from the encoded matrix for different ligands.

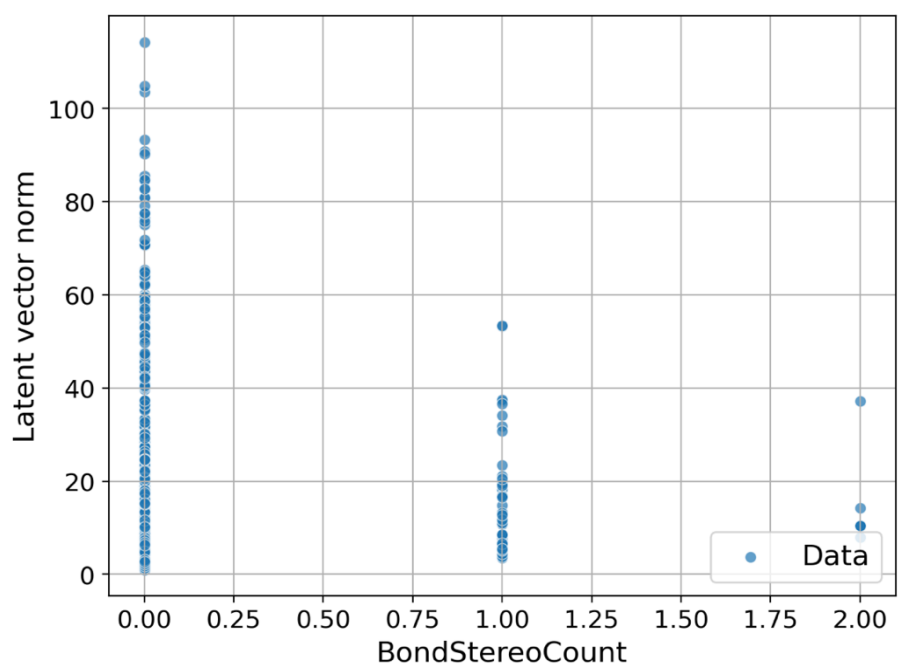

Figure S7: Here we plot the number of bonds with planar (sp2) stereo together with the $\lambda_M$ value from the encoded matrix for different ligands.

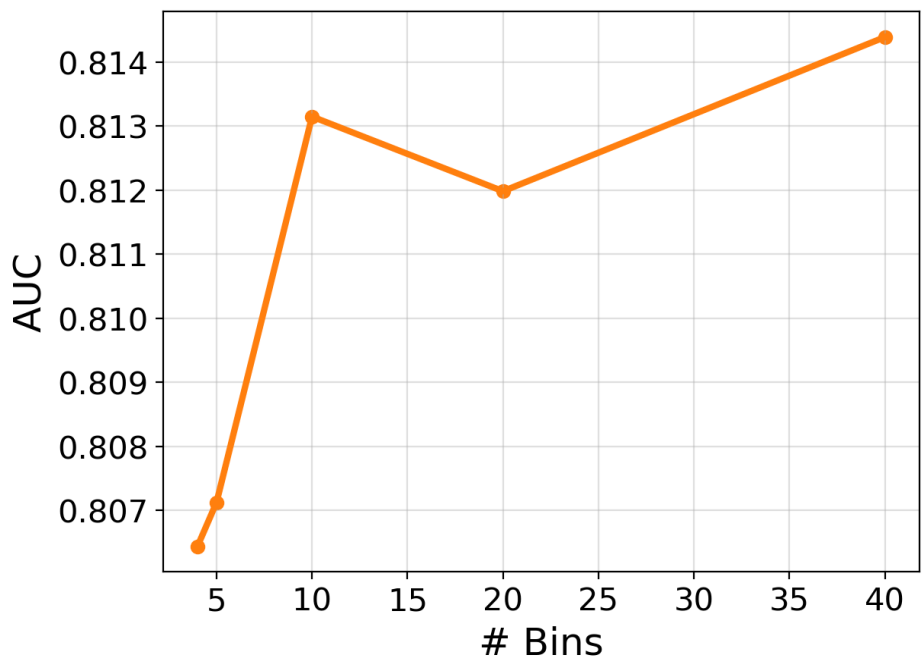

Figure S8: The AUC scores for the predictions of the random forest on the validation set of DUD-E as a function of the number of equi-spaced bins used to transform the latent vectors into an encoding matrix. We observe a small improvement inn the AUC as the number of bins increases. However, all values are above an AUC of 0.8 meaning that for any choice of the number of bins the representation is good enough to discriminate active ligands from decoys in the DUD-E dataset.

# DUD-E

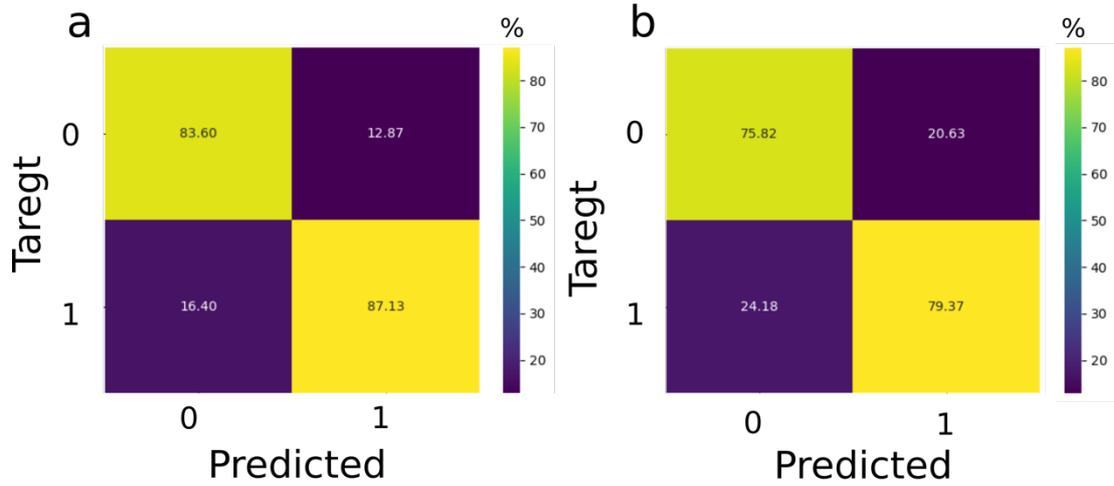

# MUV

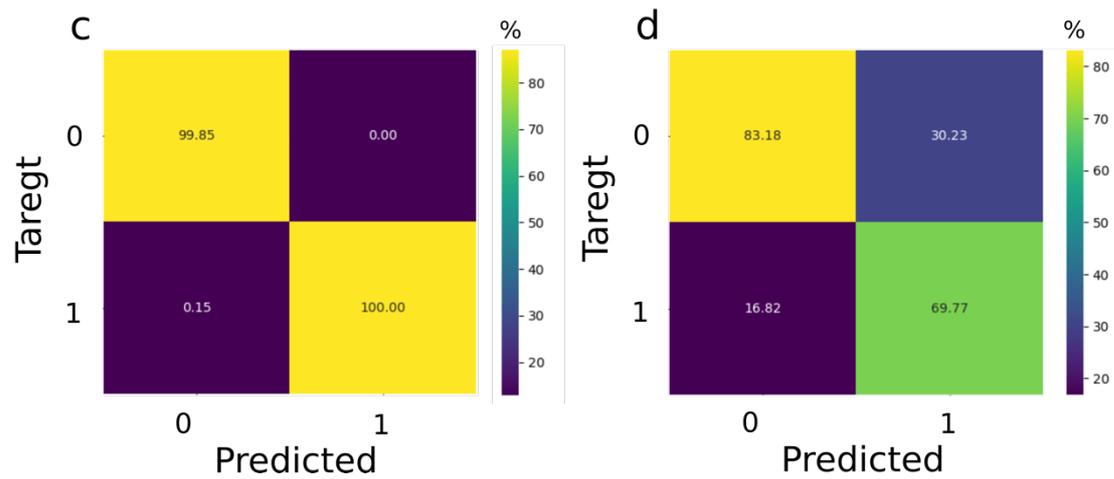

Figure S9: Confusion matrices obtained from two support vector machines (SVM) separately trained and validated on the DUD-E dataset (a and b) and the MUV dataset (c and d). a and c are obtained from the training set, b and d from the test set. The support vector machines perform slightly worse than Random Forests (with an AUC of 0.78 for DUD-E, and 0.77 for MUV).

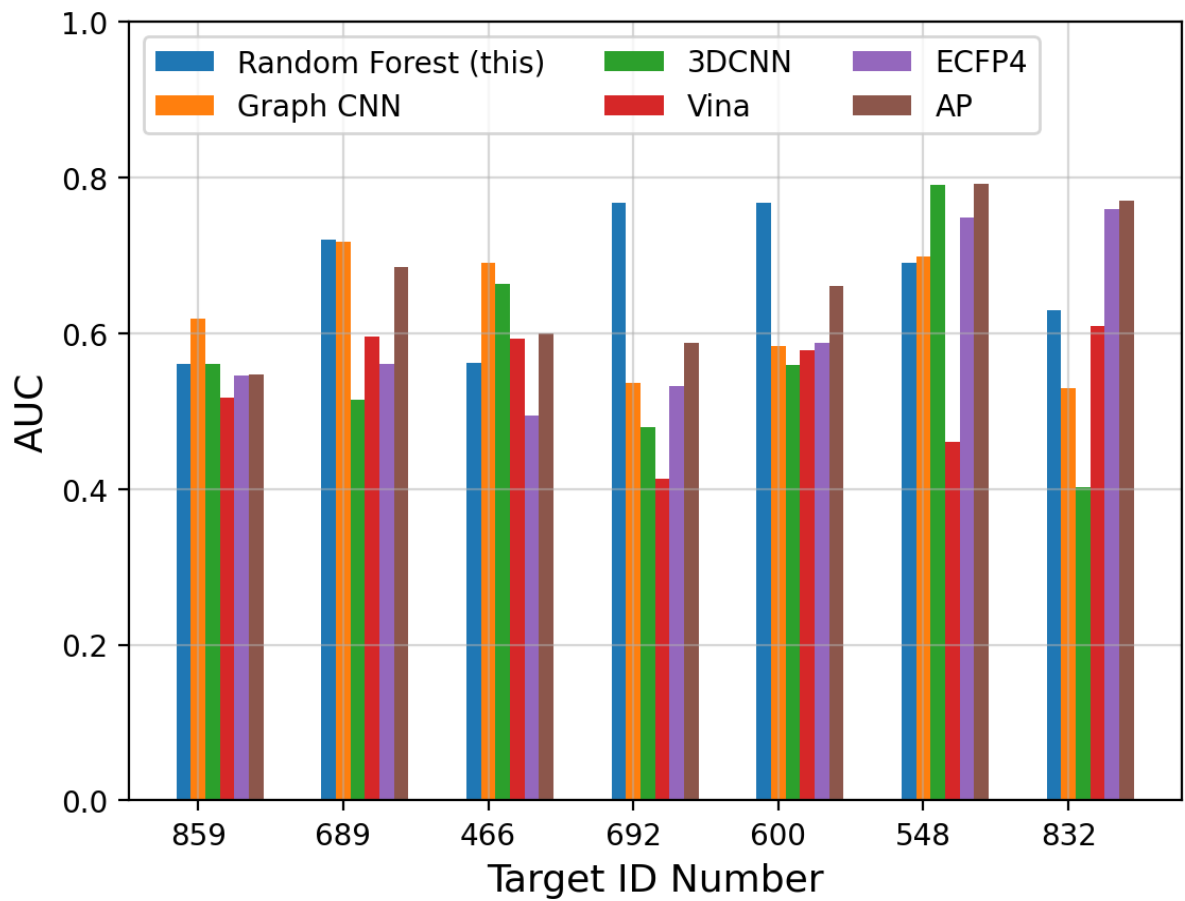

Figure S10: Comparison of AUC scores between the Random Forest (this), Graph CNN, 3DCNN, Vina, ECFP4 and AP on the different pockets of the MUV dataset.